\documentclass[10pt, conference, letterpaper]{IEEEtran}
\pdfoutput=1
\IEEEoverridecommandlockouts

\usepackage[english]{babel}
\usepackage[utf8]{inputenc}
\usepackage[T1]{fontenc}

\usepackage{amssymb}
\usepackage{amsmath}
\usepackage{graphicx}
\graphicspath{{imgs/}{figures/}{plots/}}
\usepackage[scaled]{beramono}
\usepackage[numbers]{natbib}
\usepackage{paralist}
\usepackage{textcomp}

\usepackage{algorithm}
\usepackage[noend]{algpseudocode}
\usepackage[inline]{enumitem}

\usepackage{array}
\usepackage{subfig}
\usepackage{booktabs}
\usepackage{datetime}
\usepackage{caption}
\usepackage[binary-units]{siunitx}
\usepackage[capitalize,noabbrev]{cleveref}

\usepackage{hyphenat}
\newcommand{\B}[1]{\mathbf{#1}} 
\newcommand{\keywords}[1]{}

\newtheorem{definition}{Definition}

\author{\IEEEauthorblockN{Fabien Geyer}
	\IEEEauthorblockA{Technical University of Munich | Airbus CRT \\
		Munich, Germany
	}
	\and
	\IEEEauthorblockN{Steffen Bondorf}
	\IEEEauthorblockA{Faculty of Mathematics, Center of Computer Science \\
		Ruhr University Bochum, Germany
}}

\begin{document}
\title{On the Robustness of Deep Learning-predicted Contention Models for Network Calculus}

\IEEEoverridecommandlockouts
\IEEEpubid{\makebox[\columnwidth]{978-1-7281-8086-1/20/\$31.00~\copyright2020 IEEE \hfill} \hspace{\columnsep}\makebox[\columnwidth]{ }}

\maketitle

\IEEEpubidadjcol
%

\begin{abstract}
The network calculus (NC) analysis takes a simple model consisting of a network of schedulers and data flows crossing them.
A number of analysis ``building blocks'' can then be applied to capture the model without imposing pessimistic assumptions like self-contention on tandems of servers.
Yet, adding pessimism cannot always be avoided.
To compute the best bound on a single flow's end-to-end delay thus boils down to finding the least pessimistic contention models for all tandems of schedulers in the network -- 
and an exhaustive search can easily become a very resource intensive task.
The literature proposes a promising solution to this dilemma: a heuristic making use of machine learning (ML) predictions inside the NC analysis.

While results of this work were promising in terms of delay bound quality and computational effort, there is little to no insight on when a prediction is made or if the trained algorithm can achieve similarly striking results in networks vastly differing from its training data.
In this paper, we address these pending questions.
We evaluate the influence of the training data and its features on accuracy, impact and scalability.
Additionally, we contribute an extension of the method by predicting the best $n$ contention model alternatives in order to achieve increased robustness for its application outside the training data.
Our numerical evaluation shows that good accuracy can still be achieved on large networks although we restrict the training to networks that are two orders of magnitude smaller.

\keywords{Network Calculus \and Graph Neural Network \and Robustness.}
\end{abstract}


\section{Introduction}
\label{sec:introduction}

Deterministic bounds on the end-to-end delay are strictly required in many application areas.
Prime examples are data networks in avionics and the automotive industry that are shared between multiple distributed x-by-wire applications~\cite{Geyer2016} 
as well as safety-critical (factory) systems~\cite{,Amari2016,Finzi2018,Finzi2019,Zhang2019}.

Network Calculus (NC) is a versatile framework for the derivation of such bounds.
The NC literature provides modeling and analysis tooling such that all steps towards derivation of delay bounds can be taken.
There exist results on system modeling, ranging from generic behavior like FIFO~\cite{Fidler2003,Bisti2012,Bouillard2015}, non-FIFO~\cite{Rizzo2005,Schmitt2011} or unknown~\cite{LeBoudec2001} to modern technologies such as IEEE Audio/Video Bridging (AVB) and Time-Sensitive Networking (TSN)~\cite{Queck2012,DeAzua2014,LeBoudec2018,Daigmorte2018,Zhang2019}.
Behavior of such queues, schedulers, shapers etc. are modeled as servers that, in turn, are connected to form a network, the so-called server graph~\cite{Bondorf2014,Cattelan2017}.

The server graph carries data flows, exactly one of which will be the designated flow of interest (foi) whose delay is bounded by the NC analysis.
For this network analysis step, results have been created to capture the modeled system behavior as closely as possible.
For example, on tandems of work-conserving servers, neither the worst-case burstiness of the foi nor its cross-flows should impact the delay bound computation more than once -- i.e., contention for the forwarding resource should not be assumed more pessimistically by the analysis than actually modeled by the server graph. 
These two core properties of NC are known as pay bursts only once (PBOO)~\cite{LeBoudec2001,Fidler2003,Chen2015,Tang2019} and pay multiplexing only once (PMOO)~\cite{Schmitt2008b,Bouillard2008b}, respectively.
Other such refinements try to reduce the amount of mutually exclusive contention assumptions for multiple flows at shared servers (pay segregation only once, PSOO)~\cite{Bondorf2016c}, paying for multiplexing in ring networks less often (pay multiplexing only at convergence points, PMOC)~\cite{Amari2017}, capping the worst-case burstiness with a server's queue length~\cite{Bondorf2016b} or using an entirely different alternative to compute bounds on the arrivals of cross-flows~\cite{Bondorf2018}.
Some of these results are mutually exclusive, e.g., the different arrival bounding method is based on violating the PSOO property.

Correctly capturing these restrictions on realistic worst-case contention in the given model of connected servers does not only help to improve the computed delay bound.
It also allows to rank different networks more accurately by not discriminating an alternative that features a design element NC can only consider by pessimistic over-approximation.
This allows NC to be used to compare existing network designs to newly proposed ones~\cite{Amari2016}.
However, there is not a single-best NC analysis\footnote{In this paper, we restrict our presentation to the algebraic analysis methods. The optimization analyses in~\cite{Bouillard2010,Bouillard2015} are indeed best w.r.t. to delay bounds but as shown in~\cite{Bondorf2017a,Bouillard2015}, they tend to become computationally infeasible.}.
In this paper, we focus on networks where there is no knowledge about the multiplexing behavior of flows.
This assumption is called arbitrary (or blind) multiplexing.
The best delay bound for a flow crossing a cycle-free network of arbitrary multiplexing servers is computed by a specific combination of the properties, the ``building blocks'', mentioned above.
The exhaustive search for this combination has been improved such that it becomes feasible to execute\footnote{Moreover, its bounds are very close to the optimization approach of~\cite{Bouillard2010}.} but it still tends to scale superlinearly with the network size~\cite{Bondorf2017a}.
This search-based analysis is called tandem matching analysis (TMA).

Based on TMA, DeepTMA~\cite{Geyer2019} was recently proposed to alleviate this search-induced problem.
DeepTMA is a fast heuristic based on deep learning (DL) that replaces the expensive search with a prediction of the best combination of existing building blocks, i.e., the best contention model.
While DeepTMA showed promising results towards fast and accurate NC analysis, understanding how predictions are made remains opaque such that it does not give insights in the NC analysis or the wider applicability of the method.
We aim in this paper to address those drawbacks by evaluating the influence of the dataset used in the training phase, as well as its features, on the eventual prediction accuracy.
We also contribute an extension of DeepTMA which is able to generate more than one contention model prediction, leading to an increase of the robustness of the method.

We show that DeepTMA is able to cope with scalability, namely that it can be trained on small networks and being used on much larger networks with low impact on the accuracy.
Our numerical evaluation illustrates that the relative error of DeepTMA is still below \SI{1}{\percent} on average when evaluated on networks two orders of magnitude larger than the ones used for training.
We also show that training DeepTMA on random networks leads to good applicability on more specific types of networks.
Additionally, we give insight into the importance of network features with respect to predicting a contention model.
Overall, we first demonstrate DeepTMA's robustness regarding the relation of training set to evaluated network in terms of size and shape.
Finally, we evaluate our extension of DeepTMA that proposes multiple alternative contention models.
Our evaluation shows that the robustness of DeepTMA can be increased by generating multiple contention models, leading to a decrease of the error by a factor of~2.

The remainder of the paper is organized as follows:
First, we review related work in \Cref{sec:relatedwork}.
\Cref{sec:graphnn} introduces Graph Neural Networks (GNNs) and their combination with Network Calculus.
In \Cref{sec:robustness}, we present our extension of DeepTMA and the generation of a dataset to learn from.
A numerical evaluation of the robustness of DeepTMA is performed in \cref{sec:evaluation}.
Finally, \cref{sec:conclusion} concludes.

\section{Related work}
\label{sec:relatedwork}

Research on combining machine learning with formal methods has been found in a variety of applications, e.g., in theorem proving, model-checking or in SAT-SMT problems.
In the following, we aim to provide a focused depiction of efforts that are interesting and related to our work.
Namely, the performance in networks and Graph Neural Networks (GNNs).
A more comprehensive survey on machine learning-assisted formal methods can be found in~\cite{Amrani2018}.

GNNs were first introduced in~\cite{Gori2005,Scarselli2009},
a concept subsequently refined in recent works.
Message-passing neural network were introduced in~\cite{Gilmer2017}, with the goal of unifying various GNN and graph convolutional concepts.
\cite{Velickovic2018} formalized graph attention networks, which enables to learn edge weights of a node neighborhood.
Finally, \cite{Battaglia2018} introduced the graph networks (GN) framework, a unified formalization of many concepts applied in GNNs.

These concepts were applied to many domains where problems can be modeled as graphs: chemistry with molecule analysis~\cite{Duvenaud2015,Gilmer2017}, solving the traveling salesman problem~\cite{Prates2019}, prediction of satisfiability of SAT problems~\cite{Selsam2018}, or basic logical reasoning tasks and program verification~\cite{Li2016a}.
For computer networks, they have recently been applied to prediction of average queuing delay~\cite{Rusek2018}, different non-NC-based performance evaluations of networks~\cite{Geyer2017c,Geyer2018e,Rusek2019}, and routing \cite{Geyer2018c}.
In the realm of NC, there is surprisingly little work as of yet. 
Predating DeepTMA~\cite{Geyer2019} we base our work on, there is an effort to predict the delay bound computed by different NC analyses by using GNNs.
Each of these analyses only considers a pre-defined contention model whenever there are alternatives for a tandem.
The prediction is then used to only execute the most promising analysis~\cite{Geyer2018d}.
This was developed into DeepTMA that can provide multiple predictions per analysis.
Independent efforts aim at predicting delay bounds, too.
This work~\cite{Mai2019a,Mai2019b} uses supervised learning and benchmarks the predictions against a NC-based analysis.
Another similar goal to our work is to provide small yet controllable computation times to make the proposed analysis fit for application in design space exploration.

Regarding assessing the robustness of GNNs, \cite{Selsam2018,Prates2019} showed that GNNs can be trained on a given set of graphs while being able to extrapolate on other types or much larger graphs.
Finally, \cite{Ying2019} recently proposed an approach to explain predictions from GNNs, by reducing the input graphs to subgraphs containing a small subset of nodes which are most influential for the prediction.


\section{Background: Graph Neural Network for NC}
\label{sec:graphnn}

\begin{figure}[h!]
	\centering
	\includegraphics[width=.8\columnwidth]{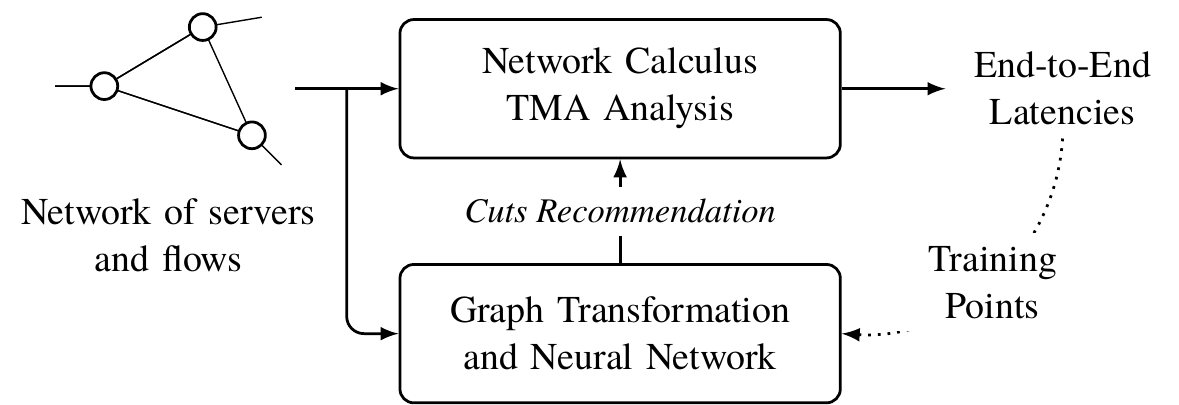}
	\caption{Overview of the proposed approach.}
	\label{fig:approach}
\end{figure}

We give a brief overview of the DeepTMA heuristic in this section.
We refer the reader to \cite{Geyer2019} for the full formulation of the method.
It is based on the concept of Graph Neural Network (GNN) introduced in~\cite{Gori2005,Scarselli2009}.
The goal of DeepTMA is to predict the best tandem decompositions, i.e., contention models, to use in TMA.
We define networks to be in the NC modeling domain and to consist of servers, crossed by flows. 
We refer to the model used in GNN as graphs. 
The main intuition is to transform the networks into graphs.
Those graph representations are then used as inputs for a neural network architecture able to process general graphs, 
which will then predict the tandem decomposition resulting in the best residual service curve.
Our approach is illustrated in \cref{fig:approach}.
Since the delay bounds are still computed using the formal network calculus analysis, they inherit its proven correctness.

\subsection{Overview of Graph Neural Networks}
\label{sec:graphnn:overview}

In this section, we detail the neural network architecture used for training neural networks on graphs, namely the family of architectures based on GNNs~\cite{Gori2005,Scarselli2009}.

Let $\mathcal{G} = (\mathcal{V}, \mathcal{E})$ be an undirected graph with nodes $v \in \mathcal{V}$ and edges $(v, u) \in \mathcal{E}$.
Let $\B{i}_v$ and $\B{o}_v$ represent respectively the input features and output values for node $v$.
The concept behind GNNs is called \emph{message passing}, where hidden states of nodes $\B{h}_v$ (i.e. vectors of real numbers) are iteratively passed between neighboring nodes.

At each iteration of message passing, each node in the graph aggregates the hidden states of its neighbors, use this aggregate to update its own hidden state, and sends the updated state at the next iteration:
\begin{align}
\B{h}_v^{(t=0)} & = \mathit{init}\left(\B{i}_v \right) \label{eq:gnn_init} \\
\B{h}_v^{(t+1)} & = \mathit{aggr}\left( \left\{ \B{h}_{u}^{(t)} \; \middle| \; u \in \textsc{Nbr}(v) \right\} \right) \label{eq:gnn_f_def}
\end{align}
with $\B{h}_v^{(t)}$ representing the hidden state of node $v$ at iteration $t$, $\mathit{aggr}$ a function which aggregates the set of hidden states of the neighboring nodes $\textsc{Nbr}(v)$ of $v$, and $\mathit{init}$ a function for initializing the hidden states based on the input features.
Those hidden states are propagated throughout the graph using multiple iterations of \cref{eq:gnn_f_def} until a fixed point is found.
The final hidden state is then used for predicting properties about nodes:
\begin{equation}
\B{o}_v = \mathit{out}\left( \B{h}_v^{(t \to \infty)} \right) \label{eq:gnn_output}
\end{equation}
with $\mathit{out}$ a function transforming $\B{h}$ to the target values.

In GNNs, the aggregation of hidden states corresponds to their sum, and the $\mathit{aggr}$ and $\mathit{out}$ functions are feed-forward neural networks (FFNN) such that:
\begin{equation}
\label{eq:gnn_f_sum}
\resizebox{.91\columnwidth}{!}{%
$\B{h}_v^{(t+1)} = \mathit{aggr}\left(\left\{ \B{h}_{u}^{(t)} \; \middle| \; u \in \textsc{Nbr}(v) \right\}\right)  = \mathit{aggr}\left( \sum_{u \in \textsc{Nbr}(v)}  \B{h}_{u}^{(t)} \right)$}
\end{equation}

Various extensions of GNNs have been recently proposed in the literature.
We selected Gated Graph Neural Networks (GGNN)~\cite{Li2016a} for implementing DeepTMA, extended with an attention mechanism similar to the one proposed in~\cite{Velickovic2018}.
This extension implements $\mathit{aggr}$ using a recurrent unit and unrolls \Cref{eq:gnn_f_def} for a fixed number of iterations.
This simple transformation allows for commonly found architectures and training algorithms for standard FFNNs as applied in computer vision or natural language processing.

In order to propagate the hidden states throughout the complete graph, a fixed number of iterations is done. 
This extension has been shown to outperform the original GNNs that required to run the iteration until a fixed point was found.
We refer to~\cite{Battaglia2018} for additional details on GNNs.

\subsection{Application to TMA} 
\label{sec:graphnn:graphmodel}

\begin{figure}[h!]
	\centering
	\subfloat[Network calculus server graph]{\includegraphics[width=.52\columnwidth]{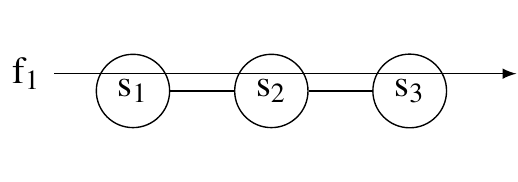}}
	\hspace{3mm}
	\subfloat[Graph representation]{\includegraphics[width=.38\columnwidth,page=2]{figures/example_network.pdf}}
	\caption{Graph representation of a sample tandem network.}
	\label{fig:example_network:graph}
\end{figure}

In order to apply the concepts described in \cref{sec:graphnn:overview} to a network calculus analysis, we transform a simple NC network (server graph with one data flow) into a graph.
\Cref{fig:example_network:graph} illustrates this transformation.

Each server is represented as a node in the graph, with edges corresponding to the network's links.
Each flow is represented as a node.
The path taken by a flow in this graph is encoded using undirected edges that connect the flow to the servers it traverses.
Since those edges do not encode the order in which those servers are traversed, so-called \emph{path ordering} nodes are added to edges between the flow node and the traversed server nodes.
This property is especially important in the TMA since the order has an impact on dependency structures and contention models.
TMA ``cuts'' the network between pairs of servers. 
The potential TMA cuts between on the path traversed by the flow of interest are represented nodes.
A cut node is connected via edges to its flow and to its pair of servers.

In addition to a categorical encoding of the node type (i.e., server, flow, path ordering or cut), the input features of each node in the graph need to comprise some NC definitions.
A comprehensive treatment of NC can be found in~\cite{LeBoudec2001}.
TMA and DeepTMA are described in~\cite{Bondorf2017a} and~\cite{Geyer2019}, respectively.
NC resource models rely on non-negative, wide-sense increasing functions
\[
\mathcal{F}_{0}\,=\,\left\{ \left.f:\mathbb{R}\rightarrow\mathbb{R}_{\infty}^{+}\;\right|\;f\left(0\right)=0,\;\forall s\le t\,:\,f(s)\!\leq\!f(t)\right\}\!,
\]
where 
$\mathbb{R}_{\infty}^{+}:=\left[0,+\infty\right)\cup\left\{+\infty\right\}$.
These functions pass through the origin.
The functions $A(t)$ and $A'(t)$ cumulatively count input and output data (in absolute time $t$) and are the basis of NC curves (interval time $d$).
The server graph crossed by flows, in short network, is annotated with the following curves, each one bounding a relevant property.

\begin{definition}[Arrival Curve]
	 \textit{Let the data arrivals of a flow over time be characterized by function \mbox{$A(t)\in\mathcal{F}_{0}$}, where \mbox{$t\in\mathbb{R}_{\infty}^{+}$}.
	An arrival curve \mbox{$\alpha(d)\in\mathcal{F}_{0}$} for $A(t)$ must then fulfill
	\[
	\forall t\,\forall d,\,0\leq d\le t\,:\,A(t)-A(t-d)\leq\alpha(d),
	\]
	i.e., it must bound the flow's data arrivals in any duration $d$.}
\end{definition}

\begin{definition}[\emph{Strict} Service Curve]
	\textit{If, during any period with backlogged data of duration~$d$, 
	a scheduler, queue, etc. with input function~$A$ guarantees an output of at least $\beta(d)\in\mathcal{F}_{0}$, 
	then it is said to offer a \emph{strict} service curve $\beta$.}
\end{definition}
\vspace{2mm}

Curves are in $\mathcal{F}_{0}$, too, and will thus not permit for data occurrences or resource availability in time intervals of duration~$0$.
They are incorporated as input features as follows:

\begin{itemize}
	\item For each server $s$, parameters of its rate-latency service curve are used where $\beta_{s}(d)=\max\left\{0,\mathit{rate}_{s}\cdot d - \mathit{latency}_{s}\right\}$: $[\mathit{rate}_{s}, \mathit{latency}_{s}]$.
	\item For each flow $f$, parameters of its token bucket arrival curve are used where $\alpha_{f}(d)= \left\{\mathit{rate}_{f}\cdot d + \mathit{burst}_{f}\right\}_{\left\{d>0\right\}}$ (i.e., $\alpha_{f}(d)=0$ for $d\leq 0$): $[\mathit{rate}_{f}, \mathit{burst}_{f}]$.
	\item For each path ordering node $p$, the hop count is encoded as an integer $\mathit{PathOrder}$.
	\item Finally, cut nodes do not have input features.
\end{itemize}
Note, that in case more complex arrival or service curve types than affine curves~\cite{Bouillard2008} are studied, those input features can be extended to represent the additional curve parameters.
Last, note that edges have no features in this graph encoding.

Since the goal of DeepTMA is to predict which tandem decomposition will result in the tightest bound, only the nodes presenting cuts have output features.
This problem is formulated as a classification problem, namely each cut node has to be classified in two classes: perform a cut between the pair of servers it is connected to or don't: $[\mathit{cut}, \overline{\mathit{cut}}]$.
The overall prediction to be fed back, i.e., the selection of one out of TMA's potential decompositions for a given foi's path, is defined by the set of all $\mathit{cut}$ classifications for this path.

\section{Increasing the Robustness of DeepTMA}
\label{sec:robustness}


\subsection{DeepTMA\textsubscript{$n$}: Generating multiple tandem decompositions}
\label{sec:graphnn:multiple}

Given a foi and a potential cut location, the output of the neural network is a probability of cutting.
This probability is generated by the neural network using the sigmoid function after its last layer.
A set of cuts is also called a ``tandem decomposition''.
In case a single tandem decomposition has to be generated, the decision of cutting is made using a threshold of \SI{50}{\percent}.

Those cut probabilities may also be used to generate multiple tandem decompositions as illustrated in \cref{algo:multipletandems}.
In case the number of all potential tandem decompositions is lower than the number of requested ones, we simply return all of them.
Otherwise, we sample the distribution of cuts in order to generate the decompositions.
We label this extension of DeepTMA as DeepTMA\textsubscript{$n$}, with $n$ the number of tandem decompositions generated.

\begin{algorithm}[h!]
	\caption{Generation of $n$ tandem decompositions for a flow traversing $L+1$ servers.}
	\label{algo:multipletandems}
	\begin{algorithmic}
		\If{$n \leq L^2$}
			\Return all combinations of cuts
		\Else
			\ForAll{$i := 1$ \textbf{to} $n$}
				\State $v \gets [c_1, \ldots, c_{L}] \sim \mathcal{U}(0, 1)^{L}$
				\State cuts\textsubscript{$i$} $\gets \mathbb{I}\left( v \leq \left[ \Pr(\mathit{cut}_{foi, 1}^{GNN}), \ldots, \Pr(\mathit{cut}_{foi, L}^{GNN}) \right] \right)$
				\State \hspace{10pt} ($\mathbb{I}$ is the indicator function)
			\EndFor
			\Return \{cuts\textsubscript{$1$}, \ldots, cuts\textsubscript{$n$}\}
		\EndIf
	\end{algorithmic}
\end{algorithm}

\subsection{Dataset generation}
\label{sec:dataset_generation}

In order to train our neural network architecture, we randomly generated a set of topologies according to three different random topology generators:
\begin{enumerate*}[label=\textit{\alph*)}]
	\item tandems or daisy-chains,
	\item trees and
	\item random server graphs following the $G(n,p)$ Erdős–Rényi model~\cite{Erdos1959}.
\end{enumerate*}
For each created server, a rate latency service curve was generated with uniformly random rate and latency parameters.
A random number of flows is generated with random source and sink servers. 
Note that in our topologies, there cannot be cyclic dependency between the flows.
For each flow, a token bucket arrival curve was generated with uniformly random burst and rate parameters.
All curve parameters were normalized to the $(0, 1]$ interval.

In total, \num{172374} different networks were generated, with a total of more than 13 million flows, and close to 260 million tandem decompositions.
Half of the networks were used for training the neural network, while the other half was used for the evaluation presented later in \Cref{sec:evaluation}.
\Cref{tab:dataset_stats} summarizes different statistics about the generated dataset.
The dataset is available online to reproduce our learning results\footnote{\texttt{https://github.com/fabgeyer/dataset-deeptma-extension}}.
Note that compared to the original dataset used for training DeepTMA \cite{Geyer2019}, this dataset contains larger networks.

\begin{table}[h!]
	\centering
	\resizebox{%
		\ifdim\width>\columnwidth
		\columnwidth
		\else
		\width
		\fi
	}{!}{%
	\begin{tabular}{@{}l|rrrr@{}}
		\toprule
		\textbf{Parameter}                & \quad\textbf{Min} & \quad\quad\textbf{Max} & \quad\textbf{Mean} & \textbf{Median} \\ \midrule
		\# of servers                     &            2 &           41 &          14.6 &              12 \\
		\# of flows                       &            3 &          203 &         101.2 &             100 \\
		\# of tandem combinations         &            2 & \num{197196} &  \num{1508.5} &             384 \\
		\# of nodes in analyzed graph     &           10 &   \num{2093} &         545.2 &             504 \\
		\# of tandem combination per flow &            2 &  \num{65536} &          19.4 &               4 \\
		\# of flows per server            &            1 &          173 &          18.1 &              10 \\ \bottomrule
	\end{tabular}}
	\caption{Statistics about the randomly generated dataset.}
	\label{tab:dataset_stats}
\end{table}

Additionally to this dataset, we also evaluate our approach on the set of networks used in \cite{Bondorf2017a}.
\Cref{tab:dataset_stats_pomacs2017} summarizes different statistics about the generated dataset.
Compared to the dataset used for training, this additional set of networks is up to two orders of magnitude larger in terms of number of servers and flows per network.
This property will be used in \cref{sec:evaluation} to investigate if our approach is able to scale to such larger networks, both in terms of accuracy and execution time, without tailoring the training set.

\begin{table}[h!]
	\centering
	\resizebox{%
		\ifdim\width>\columnwidth
		\columnwidth
		\else
		\width
		\fi
	}{!}{%
	\begin{tabular}{@{}l|rrrr@{}}
		\toprule
		\textbf{Parameter}                & \quad\textbf{Min} & \quad\quad\textbf{Max} & \quad\textbf{Mean} & \textbf{Median} \\ \midrule
		\# of servers                     &           38 &   \num{3626} &         863.0 &             693 \\
		\# of flows                       &          152 &  \num{14504} &  \num{3452.0} &      \num{2772} \\
		\# of tandem combinations         &   \num{2418} & \num{121860} & \num{24777.6} &     \num{18869} \\
		\# of nodes in analyzed graph     &   \num{1358} & \num{113162} & \num{25137.7} &     \num{19518} \\
		\# of tandem combination per flow &            2 &          512 &           7.3 &               8 \\
		\# of flows per server            &            1 &          467 &          16.4 &              12 \\ \bottomrule
	\end{tabular}}
	\caption{Statistics about the set of networks from \cite{Bondorf2017a}.}
	\label{tab:dataset_stats_pomacs2017}
\end{table}


\section{Numerical evaluation}
\label{sec:evaluation}

We evaluate in this section our extensions of DeepTMA as well as its robustness and scalability.
Via a numerical evaluation, we illustrate the tightness and execution time of DeepTMA and highlight its usability for practical use-cases.

In order to numerically evaluate and compare DeepTMA against TMA, we selected the relative error metric as our main metric for the rest of this evaluation. 
This metric measures the relative difference of DeepTMA against TMA with respect to the end-to-end delay bound, and is defined for flow $f_i$ as:
\begin{equation}
	\mathit{RelErr}_{f_i} = \frac{\mathit{Delay}_{f_i}^{\mathit{DeepTMA}} - \mathit{Delay}_{f_i}^{\mathit{TMA}}}{\mathit{Delay}_{f_i}^{\mathit{TMA}}}
\end{equation}

\subsection{Impact of training network sizes on error}

We first investigate the scalability of DeepTMA by evaluating the impact of the training dataset on the accuracy of the method.
We trained here two additional instances of the deep-learning part of DeepTMA. Each has a different restriction on the maximum amount of flows in the networks to be included in the training set, namely 50 and 100.

\begin{figure}[h!]
	\centering
	\includegraphics[width=.95\columnwidth]{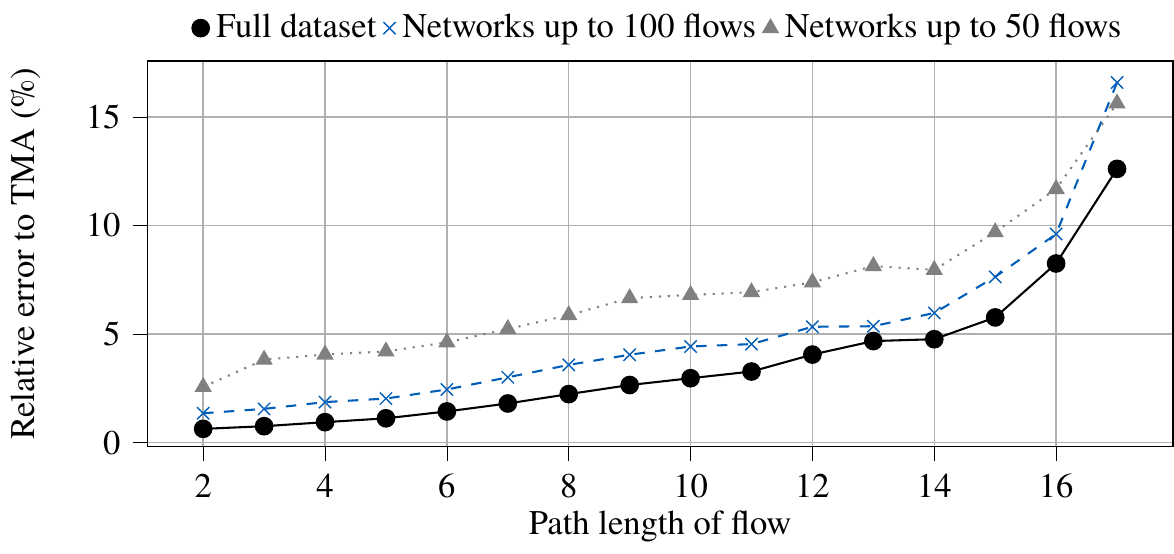}
	\caption{Impact of training size on relative error of DeepTMA.}
	\label{fig:rel_err_deeptma_dataset_size}
\end{figure}

\Cref{fig:rel_err_deeptma_dataset_size} illustrates the error of those additional instances of DeepTMA compared to the one trained on the full dataset.
There is an evident trend that a smaller training set size as imposed by our restriction generally leads to an increasing relative error.
However, there is one exception at path length 17.
This illustrates that the ``quality'' of the training set can be more important than its size.
We provide a closer look at network type and features as potential impact factors for the training set quality in the \cref{sec:eval_net-type,sec:eval_net-features}.

\begin{figure}[h!]
	\centering
	\includegraphics[width=.95\columnwidth]{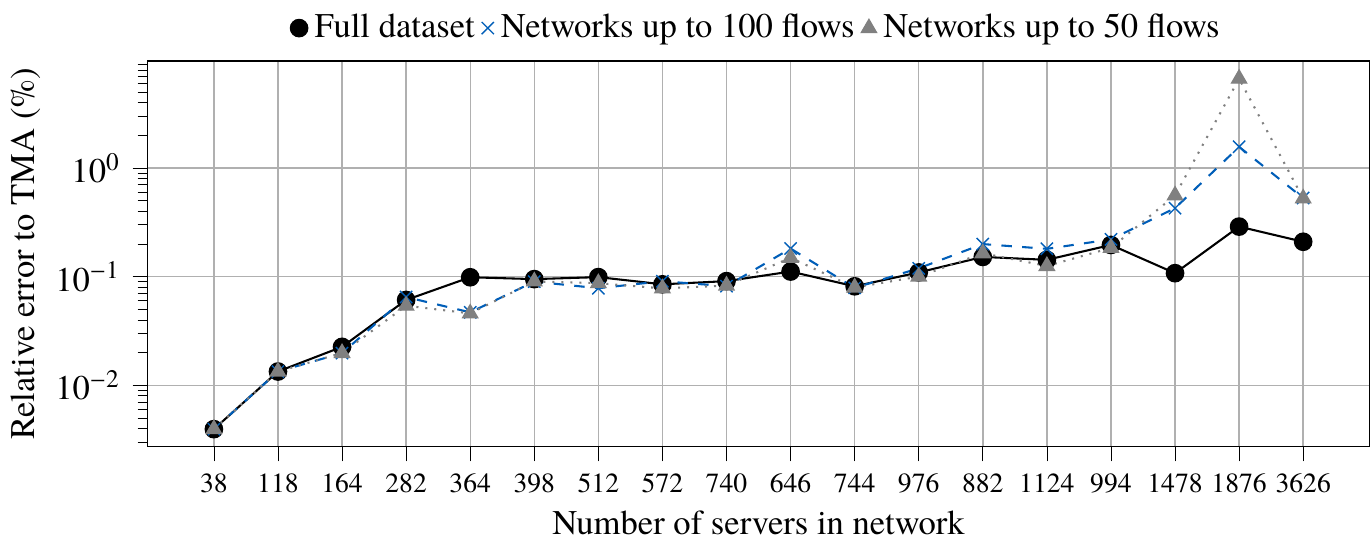}
	\caption{Impact of training size on the set of networks from \cite{Bondorf2017a}.}
	\label{fig:rel_err_pomacs2017_deeptma_trainsize}
\end{figure}

Similarly, \cref{fig:rel_err_pomacs2017_deeptma_trainsize} illustrates the impact of training dataset on the set of networks from \cite{Bondorf2017a}.
The difference between the different variants of DeepTMA is minimal except on the large networks. 
This indicates that DeepTMA is still able to scale, even when trained on much smaller networks.
Moreover, we can see small datasets outperforming the full set again in some cases.

\subsection{Influence of network type used for training}
\label{sec:eval_net-type}

We examine in this section the impact of the network types used for training DeepTMA on its accuracy.
As explained in \cref{sec:dataset_generation}, three different types of networks were generated, namely
\begin{enumerate*}[label=\textit{\alph*)}]
	\item tandems,
	\item trees and
	\item random server graphs based on the $G(n,p)$ Erdős–Rényi model.
\end{enumerate*}

\begin{figure}[h!]
	\centering
	\includegraphics[width=.95\columnwidth]{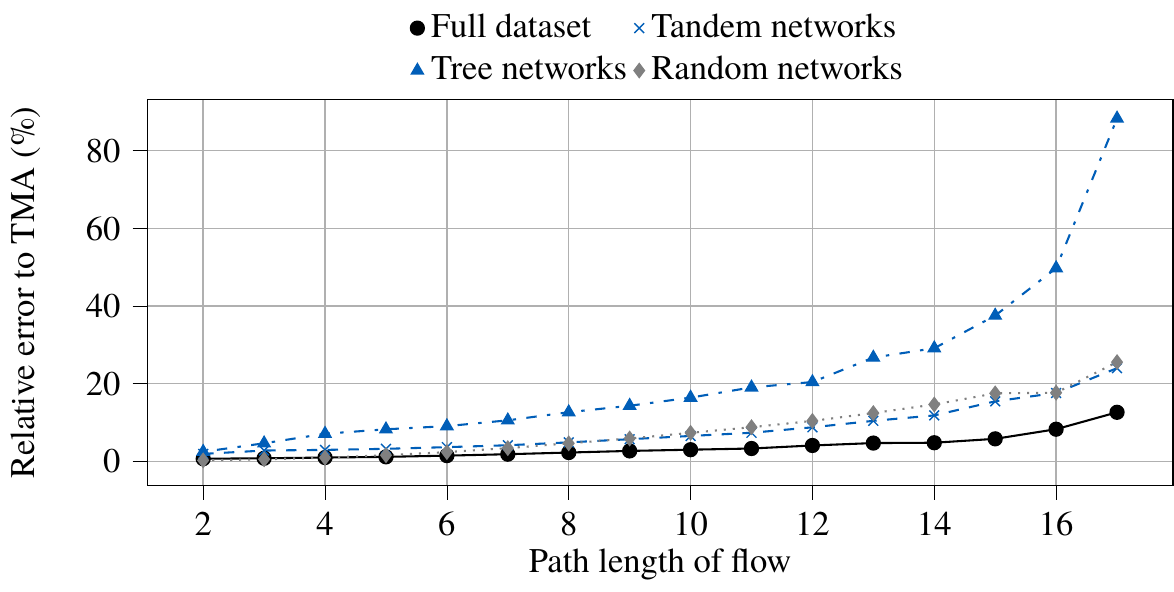}
	\caption{Impact of network types used for training on relative error of DeepTMA.}
	\label{fig:rel_err_deeptma_dataset_type}
\end{figure}

We evaluate the ability of DeepTMA to extrapolate on other networks by training three different variants for DeepTMA, each on one type of networks.
Results are presented in \cref{fig:rel_err_deeptma_dataset_type}.
Compared DeepTMA trained on the full dataset, training only on tandem or tree networks leads to good ability at extrapolating on other types of networks.
Surprisingly, tree network-based training dataset is outperformed by the tandem-based one that, in turn, is very competitive with the random server graphs.

\subsection{Importance of features and locality}
\label{sec:eval_net-features}

In order to better understand the importance of the input features used in DeepTMA, we assess each feature's importance following the permutation-based importance measure \cite{Breiman2001,Fisher2018}.
For each input feature presented in \cref{sec:graphnn:graphmodel}, we randomize it by randomly permuting its values in the training set, and assess the impact it has on the accuracy of the predictions.
We define the importance metric as:
\begin{equation}
	\resizebox{.91\columnwidth}{!}{%
	$\mathit{Importance}(\mathit{Feature}) = \frac{1}{|\mathcal{F}|} \sum_{f_i \in \mathcal{F}} \left( \mathit{RelErr}_{f_i}^{\mathit{Feature}} - \mathit{RelErr}_{f_i}^{\mathit{Baseline}} \right)$} \label{eq:featureimportance}
\end{equation}
with the baseline corresponding to DeepTMA without any feature permutation.
With this evaluation, we assess how much the GNN model relies on a given feature of interest for making its prediction.

\begin{figure}[h!]
	\centering
	\includegraphics[width=.77\columnwidth]{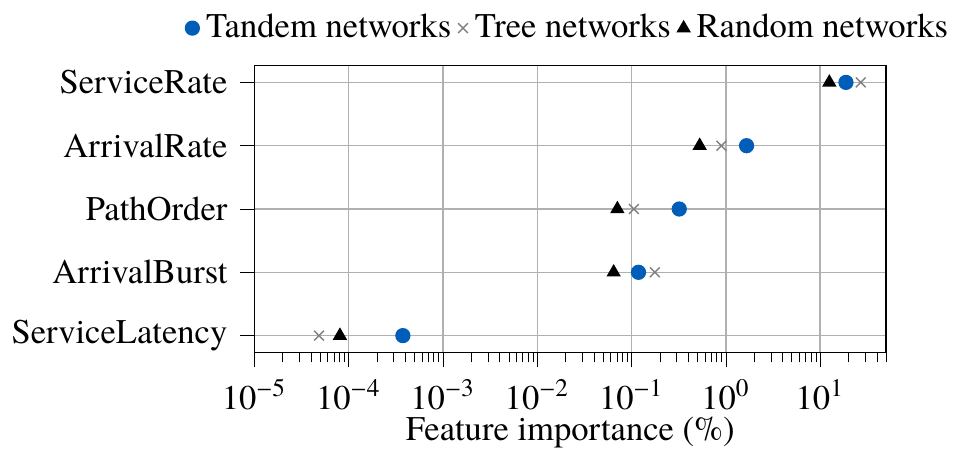}
	\\
	\includegraphics[width=.63\columnwidth]{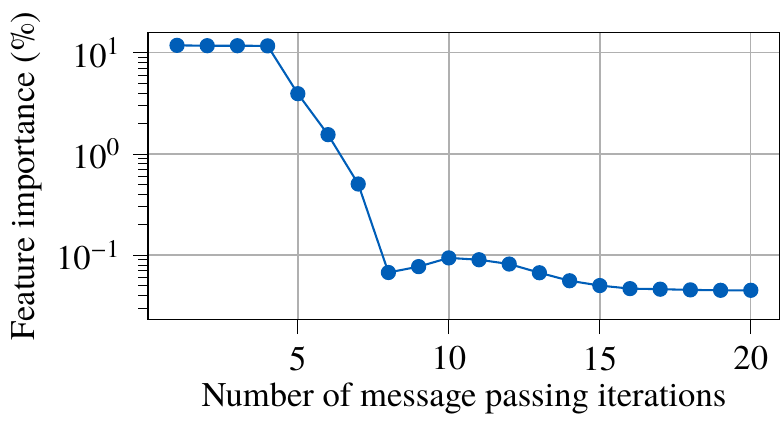}
	\caption{(above) Feature importance for DeepTMA and (below) impact of number of GNN message passing iterations.}
	\label{fig:importance}
\end{figure}

Features importance are presented in \cref{fig:importance}(above).
The service rate of the servers in the network have the largest influence on the final decision of cutting.
Such behavior confirms an existing result of NC, which is known to be sensitive to service rates.
The remaining features appear to have less importance on the cut prediction.
Interesting from the NC perspective is the observation that the order of servers (PathOrder) has a percental importance two orders of magnitude lower than the service rate.
In combination, these two features constitute the very reason for TMA (and optimization-based analyses, see~\cite{Schmitt2008}) to outperform the previous NC analyses.

We also assess the importance of other flows and other servers on a cut.
We perform this by assessing the number of iterations of message passing (i.e., \cref{eq:gnn_f_def}) and the impact it has on the relative error.
As for feature importance, we compare the results according to \cref{eq:featureimportance}.
Results are presented in \cref{fig:importance}(below).
The first 4 loop iterations appear to have the largest influence on the cut decision, meaning that the cut decision is mainly based on information from servers close to the cut.
We notice that the importance drops sharply after 5 iterations, and converges after 15 iterations.
This indicates that servers and flows farther away from the cut decision are less relevant to the cut decision~--
an insight to potential further improvement of DeepTMA's tradeoff between computational effort and relative error.

\subsection{Evaluation of DeepTMA\textsubscript{$n$}}

\begin{figure}[h!]
	\centering
	\subfloat[DeepTMA]{\includegraphics[width=.5\columnwidth]{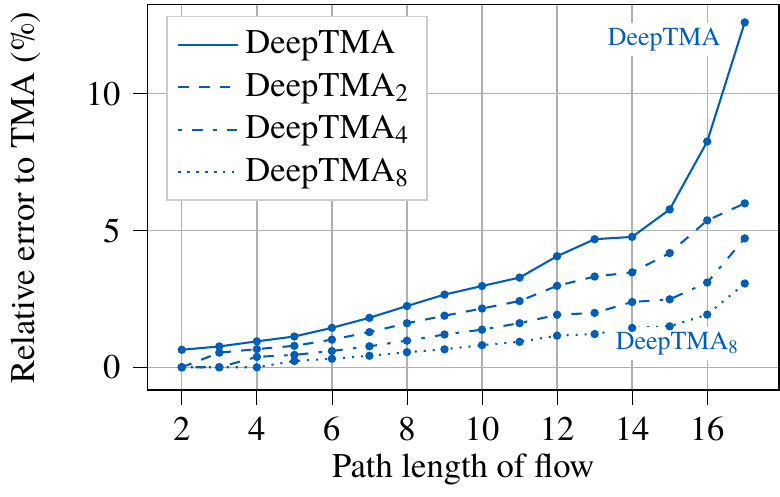}}
	\subfloat[Random heuristic]{\includegraphics[width=.5\columnwidth]{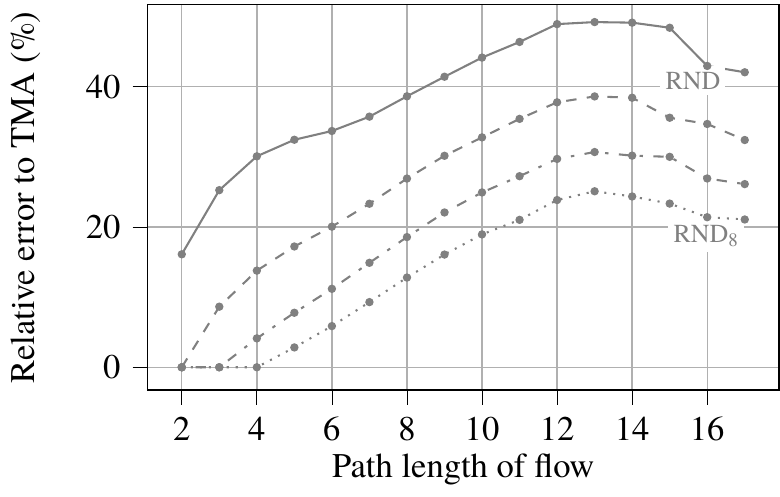}}
	\caption{Evaluation of DeepTMA\textsubscript{n} and random heuristic.}
	\label{fig:rel_err_deeptma_ns}
\end{figure}

We now start focusing on actively improving robustness and evaluate our extension of DeepTMA defined in \cref{sec:graphnn:multiple}. 
It enables DeepTMA to generate more than one tandem decomposition.
Results are presented in \cref{fig:rel_err_deeptma_ns}(a), where the subscript $n$ denotes the number of tandem decompositions generated by DeepTMA\textsubscript{$n$}.
To benchmark DeepTMA\textsubscript{$n$}, we depict the performance of a random heuristic in \cref{fig:rel_err_deeptma_ns}(b).

As expected, the generation of more than one tandem decomposition results in a decrease of the error.
Already with DeepTMA\textsubscript{2}, the error is reduced by a factor of 2 on the larger networks.
This illustrates that the robustness of DeepTMA can be increased by using \cref{algo:multipletandems},
even at the smallest additional computational cost of going to DeepTMA\textsubscript{2}.
The random heuristic performs considerably worse in all aspects evaluated above.

\subsection{Scalability on large networks}

We evaluate in this section the robustness of DeepTMA and DeepTMA\textsubscript{n} with respect to scalability.
The networks from \cite{Bondorf2017a} are evaluated here, since those networks are almost two orders of magnitude larger than the networks used for training the DeepTMA GNN, as shown in \cref{tab:dataset_stats,tab:dataset_stats_pomacs2017}.

\begin{figure}[h!]
	\centering
	\includegraphics[width=.95\columnwidth]{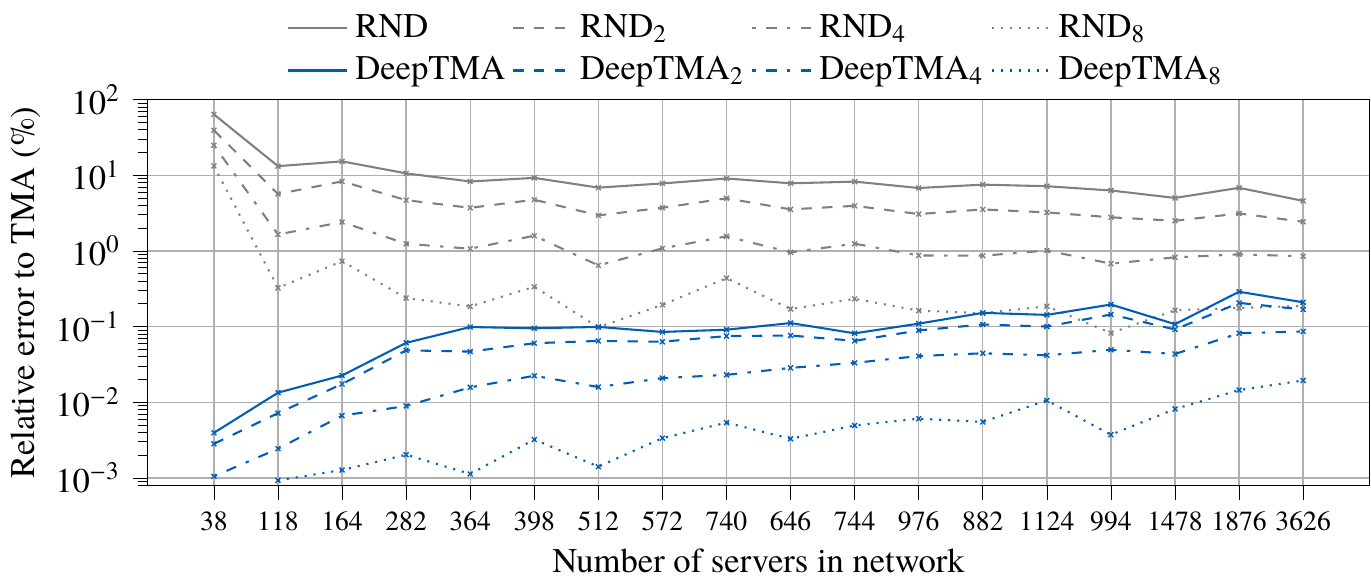}
	\caption{Evaluation of DeepTMA\textsubscript{n} on the set of networks from \cite{Bondorf2017a}.}
	\label{fig:rel_err_pomacs2017_deeptman_vs_rnd}
\end{figure}

\Cref{fig:rel_err_pomacs2017_deeptman_vs_rnd} illustrates the relative error of DeepTMA and DeepTMA\textsubscript{n} compared to a random heuristic which selects the tandem decompositions randomly.
The family of DeepTMAs achieve relative errors that are two orders of magnitudes smaller than the random heuristics, resulting in better end-to-end delay bound tightness w.r.t. the exhaustive TMA.

Although DeepTMA wasn't trained on such large networks, the relative error still stays below \SI{.3}{\percent} even on the larger networks.
DeepTMA\textsubscript{8} is even able to reach relative errors below \SI{.02}{\percent}, indicating a good ability to scale.

\section{Conclusion}
\label{sec:conclusion}

We contributed in this paper an extension of DeepTMA for generating multiple tandem decomposition predictions and a comprehensive assessment of its robustness in terms of scalability and impact of training data on its accuracy.
We also provided some insights on which feature is important for making a prediction.

Via a numerical evaluation we showed that DeepTMA can be trained on small networks and still provide good accuracy on much larger networks, up to two order of magnitude larger in terms of number of servers and flows.
We also showed that the network type used for training can have a large impact on the accuracy of the prediction made by the GNN.
Nevertheless, we showed that training DeepTMA on randomly generated networks can still lead to good accuracy, suggesting that tailoring the training data to more realistic use-cases might not be necessary for application on real networks.

Those new results indicate that DeepTMA is able to generalize tandem decomposition rules from small random networks which can also be applied to larger networks, at a low execution time cost.
Finally, we also proposed an extension of DeepTMA which is able to generate multiple predictions, decreasing the prediction error by a factor of two.

{
\footnotesize
\yyyymmdddate
\bibliographystyle{IEEEtran}
\bibliography{biblio}

\providecommand{\noopsort}[1]{}
\begin{thebibliography}{10}
\providecommand{\url}[1]{#1}
\csname url@samestyle\endcsname
\providecommand{\newblock}{\relax}
\providecommand{\bibinfo}[2]{#2}
\providecommand{\BIBentrySTDinterwordspacing}{\spaceskip=0pt\relax}
\providecommand{\BIBentryALTinterwordstretchfactor}{4}
\providecommand{\BIBentryALTinterwordspacing}{\spaceskip=\fontdimen2\font plus
\BIBentryALTinterwordstretchfactor\fontdimen3\font minus
  \fontdimen4\font\relax}
\providecommand{\BIBforeignlanguage}[2]{{%
\expandafter\ifx\csname l@#1\endcsname\relax
\typeout{** WARNING: IEEEtran.bst: No hyphenation pattern has been}%
\typeout{** loaded for the language `#1'. Using the pattern for}%
\typeout{** the default language instead.}%
\else
\language=\csname l@#1\endcsname
\fi
#2}}
\providecommand{\BIBdecl}{\relax}
\BIBdecl

\bibitem{Geyer2016}
F.~Geyer and G.~Carle, ``Network engineering for real-time networks: comparison
  of automotive and aeronautic industries approaches,'' \emph{IEEE Commun.
  Mag.}, vol.~54, no.~2, pp. 106--112, 2016.

\bibitem{Amari2016}
A.~Amari, A.~Mifdaoui, F.~Frances, and J.~Lacan, ``Worst-case timing analysis
  of {AeroRing} -- a full duplex ethernet ring for safety-critical avionics,''
  in \emph{Proc. of IEEE WFCS}, 2016.

\bibitem{Finzi2018}
A.~{Finzi}, A.~{Mifdaoui}, F.~{Frances}, and E.~{Lochin}, ``Incorporating
  {TSN/BLS} in {AFDX} for mixed-criticality applications: Model and timing
  analysis,'' in \emph{Proc. of IEEE WFCS}, 2018.

\bibitem{Finzi2019}
A.~{Finzi} and S.~S. {Craciunas}, ``Integration of {SMT}-based scheduling with
  rc network calculus analysis in {TTEthernet} networks,'' in \emph{Proc. of
  IEEE ETFA}, 2019.

\bibitem{Zhang2019}
J.~Zhang, L.~Chen, T.~Wang, and X.~Wang, ``Analysis of {TSN} for industrial
  automation based on network calculus,'' in \emph{Proc. of IEEE ETFA}, 2019.

\bibitem{Fidler2003}
M.~Fidler, ``Extending the network calculus pay bursts only once principle to
  aggregate scheduling,'' in \emph{Proc. of QoS-IP}, 2003.

\bibitem{Bisti2012}
L.~Bisti, L.~Lenzini, E.~Mingozzi, and G.~Stea, ``Numerical analysis of
  worst-case end-to-end delay bounds in fifo tandem networks,'' \emph{Real-Time
  Syst.}, 2012.

\bibitem{Bouillard2015}
A.~Bouillard and G.~Stea, ``Exact worst-case delay in {FIFO}-multiplexing
  feed-forward networks,'' \emph{IEEE/ACM Trans. Net.}, vol.~23, no.~5, pp.
  1387--1400, 2015.

\bibitem{Rizzo2005}
G.~Rizzo and J.-Y. {\noopsort{Boudec}}{Le Boudec}, ````pay bursts only once''
  does not hold for non-{FIFO} guaranteed rate nodes,'' \emph{Perform. Eval.},
  vol.~62, no. 1-4, 2005.

\bibitem{Schmitt2011}
J.~B. Schmitt, N.~Gollan, S.~Bondorf, and I.~Martinovic, ``Pay bursts only once
  holds for (some) non{-}{FIFO} systems,'' in \emph{Proc. of IEEE INFOCOM},
  2011.

\bibitem{LeBoudec2001}
J.-Y. {\noopsort{Boudec}}{Le Boudec} and P.~Thiran, \emph{Network Calculus: A
  Theory of Deterministic Queuing Systems for the Internet}.\hskip 1em plus
  0.5em minus 0.4em\relax Springer-Verlag, 2001.

\bibitem{Queck2012}
R.~Queck, ``Analysis of ethernet avb for automotive networks using network
  calculus,'' in \emph{Proc. of IEEE ICVES}, 2012.

\bibitem{DeAzua2014}
J.~A.~R. De~Azua and M.~Boyer, ``Complete modelling of avb in network calculus
  framework,'' in \emph{Proc. of RTNS}, 2014.

\bibitem{LeBoudec2018}
J.-Y. {\noopsort{Boudec}}{Le Boudec}, ``A theory of traffic regulators for
  deterministic networks with application to interleaved regulators,''
  \emph{IEEE/ACM Trans. Netw.}, vol.~26, no.~6, pp. 2721--2733, 2018.

\bibitem{Daigmorte2018}
H.~Daigmorte, M.~Boyer, and L.~Zhao, ``Modelling in network calculus a {TSN}
  architecture mixing time-triggered, credit based shaper and best-effort
  queues,'' 2018, working paper or preprint.

\bibitem{Bondorf2014}
S.~Bondorf and J.~B. Schmitt, ``The {DiscoDNC} v2 -- a comprehensive tool for
  deterministic network calculus,'' in \emph{Proc. of EAI ValueTools}, 2014.

\bibitem{Cattelan2017}
B.~Cattelan and S.~Bondorf, ``Iterative design space exploration for networks
  requiring performance guarantees,'' in \emph{Proc. of IEEE/AIAA DASC}, 2017.

\bibitem{Chen2015}
G.~Chen, K.~Huang, C.~Buckl, and A.~Knoll, ``Applying pay-burst-only-once
  principle for periodic power management in hard real-time pipelined
  multiprocessor systems,'' \emph{ACM Trans. Des. Autom. Electron. Syst.},
  2015.

\bibitem{Tang2019}
Y.~Tang, Y.~Jiang, X.~Jiang, and N.~Guan, ``Pay-burst-only-once in real-time
  calculus,'' in \emph{Proc. of IEEE RTCSA}, 2019.

\bibitem{Schmitt2008b}
J.~B. Schmitt, F.~A. Zdarsky, and I.~Martinovic, ``Improving performance bounds
  in feed-forward networks by paying multiplexing only once,'' in \emph{Proc.
  of GI/ITG MMB}, 2008.

\bibitem{Bouillard2008b}
A.~Bouillard, B.~Gaujal, S.~Lagrange, and {\'E}.~Thierry, ``Optimal routing for
  end-to-end guarantees using network calculus,'' \emph{Performance
  Evaluation}, 2015.

\bibitem{Bondorf2016c}
S.~Bondorf and J.~B. Schmitt, ``Should network calculus relocate? an assessment
  of current algebraic and optimization-based analyses,'' in \emph{Proc. of
  QEST}, 2016.

\bibitem{Amari2017}
A.~Amari and A.~Mifdaoui, ``Worst-case timing analysis of ring networks with
  cyclic dependencies using network calculus,'' in \emph{Proc. of IEEE RTCSA},
  2016.

\bibitem{Bondorf2016b}
S.~Bondorf and J.~B. Schmitt, ``Improving cross-traffic bounds in feed-forward
  networks -- there is a job for everyone,'' in \emph{Proc. of GI/ITG MMB \&
  DFT}, 2016.

\bibitem{Bondorf2018}
S.~Bondorf, P.~Nikolaus, and J.~B. Schmitt, ``Catching corner cases in network
  calculus -- flow segregation can improve accuracy,'' in \emph{Proc. of GI/ITG
  MMB}, 2018.

\bibitem{Bouillard2010}
A.~Bouillard, L.~Jouhet, and {\'E}.~Thierry, ``Tight performance bounds in the
  worst-case analysis of feed-forward networks,'' in \emph{Proc. of IEEE
  INFOCOM}, 2010.

\bibitem{Bondorf2017a}
S.~Bondorf, P.~Nikolaus, and J.~B. Schmitt, ``Quality and cost of deterministic
  network calculus -- design and evaluation of an accurate and fast analysis,''
  \emph{Proc. ACM Meas. Anal. Comput. Syst. (POMACS)}, vol.~1, no.~1, pp.
  16:1--16:34, 2017.

\bibitem{Geyer2019}
F.~Geyer and S.~Bondorf, ``{DeepTMA}: Predicting effective contention models
  for network calculus using graph neural networks,'' in \emph{Proc. of
  INFOCOM}, 2019.

\bibitem{Amrani2018}
M.~Amrani, L.~Lúcio, and A.~Bibal, ``{ML} + {FV} = $\heartsuit$? {A} survey on
  the application of machine learning to formal verification,'' 2018,
  arxiv:1806.03600.

\bibitem{Gori2005}
M.~Gori, G.~Monfardini, and F.~Scarselli, ``A new model for learning in graph
  domains,'' in \emph{Proc. of IEEE IJCNN}, 2005.

\bibitem{Scarselli2009}
F.~Scarselli, M.~Gori, A.~C. Tsoi, M.~Hagenbuchner, and G.~Monfardini, ``The
  graph neural network model,'' \emph{IEEE Trans. Neural Netw.}, vol.~20,
  no.~1, pp. 61--80, 2009.

\bibitem{Gilmer2017}
J.~Gilmer, S.~S. Schoenholz, P.~F. Riley, O.~Vinyals, and G.~E. Dahl, ``Neural
  message passing for quantum chemistry,'' in \emph{Proc. of NIPS}, 2017.

\bibitem{Velickovic2018}
P.~Veli{\v c}kovi{\'c}, G.~Cucurull, A.~Casanova, A.~Romero, P.~Li{\`o}, and
  Y.~Bengio, ``Graph attention networks,'' in \emph{Proc. of ICLR}, 2018.

\bibitem{Battaglia2018}
P.~W. Battaglia, J.~B. Hamrick, V.~Bapst, A.~Sanchez-Gonzalez, V.~Zambaldi,
  M.~Malinowski, A.~Tacchetti, D.~Raposo, A.~Santoro, R.~Faulkner, C.~Gulcehre,
  F.~Song, A.~Ballard, J.~Gilmer, G.~Dahl, A.~Vaswani, K.~Allen, C.~Nash,
  V.~Langston, C.~Dyer, N.~Heess, D.~Wierstra, P.~Kohli, M.~Botvinick,
  O.~Vinyals, Y.~Li, and R.~Pascanu, ``Relational inductive biases, deep
  learning, and graph networks,'' 2018, arxiv:1806.01261.

\bibitem{Duvenaud2015}
D.~K. Duvenaud, D.~Maclaurin, J.~Iparraguirre, R.~Bombarell, T.~Hirzel,
  A.~Aspuru-Guzik, and R.~P. Adams, ``Convolutional networks on graphs for
  learning molecular fingerprints,'' in \emph{Proc. of NIPS}, 2015.

\bibitem{Prates2019}
M.~Prates, P.~H.~C. Avelar, H.~Lemos, L.~C. Lamb, and M.~Y. Vardi, ``Learning
  to solve {NP}-complete problems: A graph neural network for decision {TSP},''
  \emph{Proc. of the {AAAI} Conference on Artificial Intelligence}, vol.~33,
  pp. 4731--4738, 2019.

\bibitem{Selsam2018}
D.~Selsam, M.~Lamm, B.~Bunz, P.~Liang, L.~de~Moura, and D.~L. Dill, ``Learning
  a {SAT} solver from single-bit supervision,'' 2018, arxiv:1802.03685.

\bibitem{Li2016a}
Y.~Li, D.~Tarlow, M.~Brockschmidt, and R.~Zemel, ``Gated graph sequence neural
  networks,'' in \emph{Proc. of ICLR}, 2016.

\bibitem{Rusek2018}
K.~Rusek and P.~Cholda, ``Message-passing neural networks learn little's law,''
  \emph{{IEEE} Communications Letters}, 2018.

\bibitem{Geyer2017c}
F.~Geyer, ``Performance evaluation of network topologies using graph-based deep
  learning,'' in \emph{Proc. of EAI ValueTools}, 2017.

\bibitem{Geyer2018e}
------, ``{DeepComNet}: Performance evaluation of network topologies using
  graph-based deep learning,'' \emph{Performance Evaluation}, 2018.

\bibitem{Rusek2019}
K.~Rusek, J.~Suárez-Varela, P.~Almasan, P.~Barlet-Ros, and
  A.~Cabellos-Aparicio, ``{RouteNet}: Leveraging {Graph Neural Networks} for
  network modeling and optimization in {SDN},'' 2019.

\bibitem{Geyer2018c}
F.~Geyer and G.~Carle, ``Learning and generating distributed routing protocols
  using graph-based deep learning,'' in \emph{Proc. Big-DAMA}, 2018.

\bibitem{Geyer2018d}
------, ``The case for a network calculus heuristic: Using insights from data
  for tighter bounds,'' in \emph{Proc. of NetCal}, 2018.

\bibitem{Mai2019a}
T.~L. Mai, N.~Navet, and J.~Migge, ``A hybrid machine learning and
  schedulability analysis method for the verification of {TSN} networks,'' in
  \emph{Proc. of IEEE WFCS}, 2019.

\bibitem{Mai2019b}
------, ``On the use of supervised machine learning for assessing
  schedulability: Application to ethernet {TSN},'' in \emph{Proc. of RTNS},
  2019.

\bibitem{Ying2019}
R.~Ying, D.~Bourgeois, J.~You, M.~Zitnik, and J.~Leskovec, ``Gnnexplainer:
  Generating explanations for graph neural networks,'' in \emph{Proc. of
  NeurIPS}, 2019.

\bibitem{Bouillard2008}
A.~Bouillard and {\'E}.~Thierry, ``An algorithmic toolbox for network
  calculus,'' \emph{Discrete Event Dynamic Systems}, vol.~18, no.~1, 2008.

\bibitem{Erdos1959}
P.~Erd{\H{o}}s and A.~R{\'e}nyi, ``On random graphs. i,'' \emph{Publicationes
  Mathematicae}, vol.~6, pp. 290--297, 1959.

\bibitem{Breiman2001}
L.~Breiman, ``Random forests,'' \emph{Machine Learning}, vol.~45, no.~1, pp.
  5--32, Oct 2001.

\bibitem{Fisher2018}
A.~Fisher, C.~Rudin, and F.~Dominici, ``Model class reliance: Variable
  importance measures for any machine learning model class, from the "rashomon"
  perspective,'' 2018.

\bibitem{Schmitt2008}
J.~B. Schmitt, F.~A. Zdarsky, and M.~Fidler, ``Delay bounds under arbitrary
  multiplexing: When network calculus leaves you in the lurch\ldots,'' in
  \emph{Proc. of IEEE INFOCOM}, 2008.

\end{thebibliography}
}

\end{document}